\documentclass[11pt]{article}
\usepackage{comment}
\usepackage{amsfonts,amssymb,epsfig,amsmath}
\usepackage{graphicx}
\usepackage{color}
\renewcommand{\text}[1]{#1}

\newcommand{\be}{\begin{equation}}
\newcommand{\ee}{\end{equation}}
\newcommand{\ben}{\begin{displaymath}}
\newcommand{\een}{\end{displaymath}}
\newcommand{\bea}{\begin{eqnarray}}
\newcommand{\eea}{\end{eqnarray}}
\newcommand{\bean}{\begin{eqnarray*}}
\newcommand{\eean}{\end{eqnarray*}}
\newcommand{\nn}{\nonumber \\}
\newcommand{\ba}{\begin{array}}
\newcommand{\ea}{\end{array}}
\newcommand{\bi}{\begin{itemize}}
\newcommand{\ei}{\end{itemize}}

\renewcommand{\theequation}{\arabic{section}.\arabic{equation}}
\def\theequation{\thesection.\arabic{equation}}

\def\a{\alpha}

\def\b{\beta}
\def\g{\gamma}
\def\G{\Gamma}

\def\G{\Gamma}
\def\g{\gamma}
\def\e{\epsilon}
\def\s{\sigma}
\def\e{\epsilon}

\begin{document}

\makeatletter
\renewcommand{\theequation}{\thesection.\arabic{equation}}
\@addtoreset{equation}{section}
\makeatother

\begin{titlepage}

\vfill
\begin{flushright}
KIAS-P11039
\end{flushright}

\vfill

\begin{center}
   \baselineskip=16pt
   {\Large\bf A search for $AdS_5 \times S^2$ IIB supergravity solutions \\  dual to $\mathcal{N}=2$ SCFTs }
   \vskip 2cm
      Eoin \'O Colg\'ain$^1$ \& Bogdan Stefa\'nski, jr.$^2$
       \vskip .6cm
             \begin{small}
      \textit{$^{1}$Korea Institute for Advanced Study, \\
        Seoul 130-722, Korea} \\ \vspace{2mm} \textit{
$^{2}$ Centre for Mathematical Science, City University London, \\
Northampton Square, London EC1V 0HB, UK
}
               \end{small}\\*[.6cm]
\end{center}

\vfill \begin{center} \textbf{Abstract}\end{center} \begin{quote} We
present a systematic search for Type IIB supergravity solutions whose
spacetimes include $AdS_5$ and $S^2$ factors, which would be candidate
duals to ${\cal N}=2$ four-dimensional Superconformal field theories. The
candidate solutions encode the $SU(2)$ R-symmetry geometrically on the
$S^2$ and an additional Killing vector generates the $U(1)$ R-symmetry. By
analysing the Killing spinor equations we show that no such solutions
exist. This suggests that if Type IIB backgrounds dual to ${\cal N}=2$
SCFTs exist, the $SU(2)$ R-symmetry is realised non-geometrically.
Finally, we also show that, in the context of both ${\cal N}=1$ and ${\cal
N}=2$ Type IIB backgrounds with an $AdS_5$ factor, the only candidate $U(1)$ R-symmetry
Killing vector directions are the ones that appear for generic values of
the Killing spinors;  no further Killing vectors exist for special
values of the Killing spinors.
%apart from the maximally supersymmetric
%${\cal N}=4$ spacetime.
\end{quote} \vfill

\end{titlepage}

%%%%%%%%%%%%%%%%%%%%%%%%%%%%%%%%%%%%%%%%%%%%%%%%%%%%%%%%%%%%%%%%%%%%%%%%%%%%%%%%%%%
%%%%%%%%%%%%%%%%%%%%%%%%%%%%%%%%%%%%%%%%%%%%%%%%%%%%%%%%%%%%%%%%%%%%%%%%%%%%%%%%%%%
\section{Introduction \& Summary}

Recently, there has been a renewed interest in ${\cal N}=2$ Superconformal field theories (SCFTs) coming
from a number of directions. Wilson- and 't Hooft-loop computations in these theories have been
performed using localization techniques in~\cite{Pestun:2007rz,Gomis:2011pf} and matrix
models~\cite{Passerini:2011fe}.
Localization methods of~\cite{Pestun:2007rz} have also provided the basis for the AGT
conjecture~\cite{Alday:2009aq,Wyllard:2009hg} which relates supersymmetric
quantities of four-dimensional ${\cal N}=2$ SCFTs to correlators in two-dimensional
SCFTs~\cite{Alday:2009fs,Drukker:2009id}.
Investigations of S-duality properties of general ${\cal N}=2$ SCFTs have led to a better understanding
of the moduli space of these theories~\cite{Argyres:2007cn,Cecotti:2011rv}, and in particular to a conjecture about the
existence of
families of strongly coupled ${\cal N}=2$ SCFTs without a Lagrangian
description~\cite{Gaiotto:2009we}. Novel
connections between integrable systems and ${\cal N}=2$ gauge theories have been discovered
in~\cite{Nekrasov:2009uh,Nekrasov:2009ui,Nekrasov:2009rc,Nekrasov:2010ka,Dorey:2011pa,Chen:2011sj}.

Within the context of the $AdS/CFT$ correspondence, a proposal for the supergravity duals of ${\cal
N}=2$ SCFTs has been made in M-theory~\cite{Gaiotto:2009gz} and in the Type IIA
reduction~\cite{bogdanIIA}\footnote{Singular solutions obtained via non-Abelian T-duality \cite{de la Ossa:1992vc} from $AdS_5 \times S^5$ appeared in \cite{Sfetsos:2010uq}.}. These spacetimes
were found using an approach developed for the study of the Killing spinor equations (KSEs) of
supergravity using spinor bilinears and the ansatz of~\cite{LLM}.\footnote{M2 and M5-brane probes in these backgrounds corresponding to loop and surface operators in the dual ${\cal N} = 2$ SCFT have been studied in \cite{Drukker:2009tz} and \cite{eoin2}.} Further, it has been
shown in~\cite{eoin1} that~\cite{LLM}
is indeed the most general solution with the chosen spacetime ansatz. Given the existence of these
M-theory and IIA solutions, a natural question is whether solutions dual to ${\cal N}=2$ SCFTs can be
found in Type IIB supergravity. In this paper we perform a systematic search for Type IIB supergravity
solutions that have ${\cal N}=2$ SCFT duals. We do this by considering spacetimes with $AdS_5$ and $S^2$
factors which realise the $SO(2,4)$ conformal and $SU(2)$ R-symmetries geometrically. A detailed
analysis of the KSEs and resulting bispinor relations reveals that no solutions beyond the maximally
supersymmetric $AdS_5\times S^5$ solution exist. The earlier paper~\cite{D'Hoker:2010mk} showed that no
IIB supergravity solutions with $AdS_5\times S^2\times S^1$ and $AdS_5\times S^3$ factors (with each
factor warped over a two-dimensional Riemann surface) exist. The first type of spacetime is a sub-class
of the ansatz that we take in this paper,~\footnote{We allow for the possibility of the $U(1)$
R-symmetry direction to be fibred and warped.} and our results are consistent with the lack of solutions
of this type.

At first sight, one might expect other ways of realising the $SU(2)$ R-symmetry geometrically. However,
in appendix F of~\cite{LLM} it is shown that the $U(1)$ Killing direction cannot be fibred over the
$S^2$ associated with the $SU(2)$ R-symmetry. Although this more general ansatz preserves $SU(2)$
symmetry, if the Killing spinor is charged under translations in the $U(1)$-direction, in other words,
if the $U(1)$ is an R-symmetry, then the supercharges cannot form an $SU(2)$ doublet. To see an example
of this, one might consider a IIB spacetime containing an $S^3$ squashed along its Hopf fibre. While
this background preserves an $SU(2)\times U(1)$ isometry the corresponding Killing
spinors~\cite{Compere:2008cw,Orlando:2010ay} transform as singlets of $SU(2)$.\footnote{We are greatful
to Linda Uruchurtu for an explanation of this and related discussions.} As a result, the only way to
realise the $SU(2)$ R-symmetry geometrically in a way that is consistent with the superconformal algebra
is by including a round $S^2$ factor in the metric.

Since we do not find any IIB solutions with such an $S^2$ factor beyond the maximally supersymmetric
$AdS_5\times S^5$ solution we conclude that if IIB supergravity duals do exist for generic ${\cal N}=2$ SCFTs the
$SU(2)$ R-symmetry is realised non-geometrically. {The possibility that the $SU(2)$ R-symmetry is realised non-geometrically, was raised in the context of non-conformal $\mathcal{N}=2$ SYM theory in~\cite{Gauntlett:2001ps,Bigazzi:2001aj}}. It is possible that by including sources into the
supergravity equations and realising the $SU(2)$ on the corresponding branes one may realise the ${\cal
N}=2$ superconformal algebra without an $S^2$ factor in the spacetime. Or perhaps in the context of IIB
the R-symmetry is only realised in the full string theory rather than the supergravity? A mild caveat to the above is that we take the internal space to be compact. Relaxing such a constraint, one still has the possibility of non-compact solutions with field content incorporating a constant five-form flux and non-constant harmonic axion and dilation, but only in the case when the complex three-form flux is zero~\cite{james}.

The structure of this paper runs as follows. In section~\ref{sec2} we review the general reduction of IIB supergravity used in~\cite{james} for spacetimes
with an $AdS_5$ factor. In section~\ref{sec3} we reduce further on a round $S^2$, and write down the
resulting algebraic and differential KSEs. Using these, in section~\ref{sec4} we look for potential
Sasaki-Einstein type solutions. In section~\ref{sec5} we search for general solutions preserving our
ansatz. Using some bispinor algebra we identify {\em two} putative Killing vector directions and find
the conditions necessary for each of these to correspond to a global $U(1)$ symmetry. In section~\ref{sec6}
we show that the global $U(1)$ symmetry constraints imply, that either the solution is the maximally supersymmetric
$AdS_5\times S^5$ or the Killing vectors are zero. Finally, in section~\ref{sec7} we show that when
these two Killing vectors are zero, the KSEs imply that the Killing spinors are also zero. This
completes the demonstration that no solutions of the abovementioned form exist in IIB supergravity. In
section~\ref{sec7} we also show that a similar argument holds more generally in the case of ${\cal N}=1$
$AdS_5$ backgrounds originally investigated in~\cite{james}. In that setting a unique Killing vector was
identified (it was denoted as $K_5$ in~\cite{james} and we will write it as $K_5^{GMSW}$). One may want
to ask what happens when we restrict the Killing
spinors in a way that makes the vector equal to zero. Our analysis shows that in that case too, the KSEs
imply that all the Killing spinors are zero, and hence, that no such solution exists.

%Following \cite{LLM}, if one denotes the canonical connection on $S^2$ as $A$, the Killing spinor equation on the $S^2$ %becomes
%\be
%\label{s2eq}
%[ \nabla_{\alpha} - A_{\alpha} \partial_{\chi} ] \eta = a \gamma_{\alpha} \eta,
%\ee
%where $\alpha =1,2$ labels sphere directions, $\chi$ denotes the $U(1)$-direction and $\gamma_{\alpha}$ denote the $S^2$
%gamma matrices. Taking the commutator of the above equation, one gets
%\be
%\label{s2eqb}
%[ -\tfrac{1}{2} \gamma_{12} - F_{12} \partial_{\chi} ] \eta = 2 a^2 \gamma_{12} \eta,
%\ee
%where, as usual, $F_{12} = \partial_1 A_2 - \partial_2 A_1$.
%\textbf{I have added a factor.} Now, if $A = 0$ and there is no connection, one recovers $a = \pm i/2$ and the usual Killing
%spinor equation on $S^2$. However, if $ \eta  \sim e^{i m \chi}$, where $m$ is the $U(1)$ charge, then (\ref{s2eqb}) can only
%be satisfied if $\eta$ is chiral satisfying $\gamma_{12} \eta = \pm i \eta$.%

\section{Type IIB Review}
\label{sec2}
In order to define some notation, in this section we review the construction of~\cite{james} for KSEs in Type IIB supergravity~\cite{IIBsugra} with a spacetime containing an $AdS_5$ factor. We will be essentially following the notation and conventions of~\cite{james}. Solutions of type IIB supergravity in Einstein frame, preserve supersymmetry as long as the following variations vanish
\bea
\delta \psi_M &\simeq& D_{M} \epsilon - \frac{1}{96} (\Gamma_{M}^{~P_1 P_2 P_3} G_{P_1 P_2 P_3} - 9 \Gamma^{ P_1 P_2} G_{M P_1 P_2}) \epsilon^{c}
+ \frac{i}{192} \Gamma^{P_1 P_2 P_3 P_4} F_{M P_1 P_2 P_3 P_4} \epsilon, \nn
\delta \lambda &\simeq& i \G^{M} P_{M} \e^c + \frac{i}{24} \G^{P_1 P_2 P_3}G_{P_1 P_2 P_3} \e,
\eea
where $F$ denotes the self-dual five-form flux, $G$ the complex three-form flux and $P$ the complex axion-dilaton. In terms of the conventional string-theory variables, these latter two may be further expressed as~\cite{hassan}
\bea
P &=& -i Q + \frac{1}{2} d \phi = \frac{i}{2} e^{\phi} d C^{(0)} + \frac{1}{2} d \phi, \nn
G &=& i e^{\phi/2} (\tau d B - d C^{(2)}),
\eea
where $\tau \equiv C^{(0)} + i e^{-\phi}$.  In addition, there is a manifest  $SL(2,\mathbb{R})$ action transforming the constituents of $P$ and $G$~\cite{hassan}.

One also has a local $U(1)$ invariance associated to the gauge field $Q_M$, with the spinor $\epsilon$, fields $P$ and $G$ charged with charge $\tfrac{1}{2}$,
$2$ and $1$ respectively. $D$ above denotes the covariant derivative incorporating this local $U(1)$ transformation, i.e.
\be
D_{M} \epsilon = \left( \nabla_{M} - \frac{i}{2} Q_M \right) \epsilon.
\ee

Once the supersymmetry conditions are imposed, to ensure a genuine supergravity solution, one has to guarantee that the following field equations of motion,
\bea
\label{10dfeom}
F &=& *_{10} F, \nn
D *_{10} G  &=& P \wedge *_{10} G^* + i F \wedge G, \nn
D *_{10} P &=& - \frac{1}{4} G \wedge *_{10} G,
\eea
and Einstein equation,
\bea
\label{10deinstein}
R_{MN} &=& P_M P_N^* + P_N P_M^* + \frac{1}{96} F_{M P_1 P_2 P_3 P_4} F_{N}^{~P_1 P_2 P_3 P_4} \nn
&+& \frac{1}{8} \left( G_{M}^{~P_1 P_2} G^*_{N P_1 P_2} + G_{N}^{~P_1 P_2} G^*_{M P_1 P_2} - \frac{1}{6} g_{MN} G^{P_1 P_2 P_3} G^*_{P_1 P_2 P_3}\right), \nn
\eea
are satisfied. Lastly, one also needs to impose the Bianchi identities
\bea
\label{10dbianchi}
d Q &=& -i P \wedge P^*, \nn
D P &=& 0, \nn
D G  &=& - P \wedge G^*, \nn
d F &=& \frac{i}{2} G \wedge G^{*}.
\eea
We remark that the Bianchi for $P$ is trivially satisfied.

However, not all these conditions on the geometry are independent. By examining integrability of the supersymmetry conditions, it was shown in  appendix D of~\cite{james} that for spacetimes with $AdS_5$ factors that the equations of motion are a consequence of supersymmetry. Our analysis will later make use of these equations of motion, so we recast them later in terms of the three-dimensional field content.

\section{Supersymmetry conditions in lower dimensions}
\label{sec3}

In this section we review the reduction of the supersymmetry conditions on $AdS_5$ presented
in~\cite{james}, and extend those results by further decomposing on a round $S^2$ to three-dimensions.
We begin with the ansatz used in~\cite{james} for the ten-dimensional spacetime
\bea
ds^2_{10} &=& e^{2 \Delta} [ ds^2(AdS_5) + ds^2(M_5) ], \nn
F &=& (vol_{AdS_5} + vol_{M_5}) f\,.
\eea
Above $f$ constant, and $Q, P$ and $G$ all can take values on $M_5$. The ten-dimensional MW spinors are
decomposed into Killing spinors on $AdS_5$ and spinors on $M_5$; the latter are denoted as $\xi_i$. As
was shown in~\cite{james} this ansatz led to two differential,
\begin{align}
   D_m \xi_1
      + \frac{i}{4} \left(e^{-4\Delta}f-2m\right)\gamma_m \xi_1
      + \frac{1}{8} e^{-2\Delta} G_{mnp}\gamma^{np}\xi_2
      &= 0 \label{sone}\\
   \bar{D}_m \xi_2
      -  \frac{i}{4} \left(e^{-4\Delta}f+2m\right)\gamma_m \xi_2
      + \frac{1}{8} e^{-2\Delta} G_{mnp}^*\gamma^{np}\xi_1
      &= 0 \label{stwo}
\end{align}
and four algebraic conditions,
\begin{align}
   \gamma^m\partial_m\Delta\xi_1
      - \frac{1}{48}e^{-2\Delta}\gamma^{mnp}G_{mnp}\xi_2
      - \frac{i}{4}\left(e^{-4\Delta}f-4m\right) \xi_1
      &= 0  \label{sthree}\\
   \gamma^m\partial_m\Delta\xi_2
      - \frac{1}{48}e^{-2\Delta}\gamma^{mnp}G_{mnp}^*\xi_1
      + \frac{i}{4}\left(e^{-4\Delta}f+4m\right)\xi_2
      &= 0 \label{sfour}\\
   \gamma^m P_m \xi_2
      + \frac{1}{24} e^{-2\Delta} \gamma^{mnp} G_{mnp} \xi_1
      &= 0 \label{sfive}\\
   \gamma^m P_m^* \xi_1
      + \frac{1}{24} e^{-2\Delta} \gamma^{mnp} G_{mnp}^* \xi_2,
      &= 0 \label{ssix}\,.
\end{align}

In this paper, we consider $M_5$ to contain an $S^2$ factor. Similar two-sphere decompositions have
appeared in~\cite{LLM, eoin1,ads3s2}, where the basic idea is to further split the spinors $\xi_i$ by decomposing them as
\be
\label{spinor5to3}
\xi_{i} = \chi_{+} \otimes \e_{i+} + \chi_{-} \otimes \e_{i-}\,,
\ee
where $\chi_{\pm}$ denote solutions to the Killing spinor equation on $S^2$
\be
\nabla_{\a} \chi_{\pm} = \pm \tfrac{i}{2} \s_{\a} \chi_{\pm}\,.
\ee
Explicit expressions for $\chi_\pm$ can be found in~\cite{s3kse}. Without loss of generality we take
$\chi_-
= \s_3 \chi_+$. Out of the spinors $\e_{i\pm}$ it is possible to construct a number of spinor bilinears that transform either as scalars or vectors on $M_3$. We refer the reader to the appendix for a complete list of these as well as our gamma matrix conventions.

To complete the task in the light of the LLM observation~\cite{LLM} that there should be no $U(1)$ fibre over the $S^2$, we adopt the warped-product ansatz for the bosonic sector
\bea
ds^2(M_5) &=& e^{2 B} ds^2(S^2) + ds^2(M_3), \nn
G &=& \mathcal{A} \wedge vol_{S^2} + {g} vol_{M_3}.
\eea
Furthermore, we use calligraphic notation to distinguish the $D=3$ fields $\mathcal{P}, \mathcal{Q}$ from the $D=5$ fields $P, Q$.

After following the decomposition through, one may extract the three-dimensional supersymmetry
conditions. We now have four differential,
\bea
\label{diff1}
&&0 = D_{m} \e_{1\pm} + \tfrac{i}{4}(e^{-4 \Delta} f - 2m) \s_m \e_{1\mp} + \tfrac{i}{4} e^{-2 \Delta} \left[ g \s_m \e_{2 \pm} - e^{-2B} \mathcal{A}_{m} \e_{2 \mp} \right], \\
\label{diff2}
&&0 = \bar{D}_{m} \e_{2\pm} - \tfrac{i}{4}(e^{-4 \Delta} f + 2m) \s_m \e_{2\mp} +  \tfrac{i}{4} e^{-2 \Delta} \left[ g^* \s_m \e_{1 \pm} - e^{-2B} \mathcal{A}^*_{m} \e_{1 \mp}
\right],
\eea
and twelve algebraic constraints
\bea
\label{alg1} &&0 = \left[ \pm i e^{-B} + \tfrac{i}{2} (e^{-4 \Delta} f - 2 m) \right] \e_{1\pm} + \s^{m} \partial_{m} B\e_{1\mp} - \tfrac{i}{2} e^{-2 \Delta - 2 B} \s^{m}
\mathcal{A}_{m} \e_{2 \pm}\,, \\
\label{alg2} &&0 = \left[ \pm i e^{-B} - \tfrac{i}{2} (e^{-4 \Delta} f + 2 m) \right] \e_{2\pm}  + \s^{m} \partial_m B \e_{2\mp} - \tfrac{i}{2} e^{-2 \Delta-2B} \s^{m}
\mathcal{A}^*_{m} \e_{1 \pm}\,, \\
\label{alg3} &&0 = \s^{m} \partial_{m} \Delta \e_{1 \pm} -  \tfrac{i}{8} e^{-2 \Delta} \left[ g \e_{2 \pm} - e^{-2 B} \s^{m} \mathcal{A}_{m} \e_{2 \mp} \right] -
\tfrac{i}{4} \left(e^{-4 \Delta} f - 4 m \right)\e_{1 \mp}\,, \\
\label{alg4} &&0 = \s^{m} \partial_{m} \Delta \e_{2 \pm} - \tfrac{i}{8} e^{-2 \Delta} \left[ g^* \e_{1 \pm} - e^{-2 B} \s^{m} \mathcal{A}^*_{m} \e_{1 \mp} \right]
+ \tfrac{i}{4} \left(e^{-4 \Delta} f + 4 m \right)\e_{2 \mp}\,, \\
\label{alg7} &&0 = \s^{m} \mathcal{P}_m \e_{2 \pm} + \tfrac{i}{4} e^{-2 \Delta} \left[ g \e_{1 \pm} - e^{-2 B} \s^{m} \mathcal{A}_{m} \e_{1 \mp} \right]\,, \\
\label{alg8} &&0 = \s^{m} \mathcal{P}_m^* \e_{1 \pm} + \tfrac{i}{4} e^{-2 \Delta} \left[ g^* \e_{2 \pm} - e^{-2B} \s^{m} \mathcal{A}^*_{m} \e_{2 \mp} \right]\,,
\eea
By combining some of the above algebraic constraints, one may also show that
\bea
\label{alg5} 0 && = \left[ \pm i e^{-B} - \tfrac{i}{2} e^{-4 \Delta } f + 3 i m \right] \e_{1\pm} + \s^m \partial_m (4 \Delta +B) \e_{1 \mp} - \tfrac{i}{2}
e^{-2 \Delta} g \e_{2 \mp}\,, \\
\label{alg6} 0 && = \left[ \pm i e^{-B} + \tfrac{i}{2} e^{-4 \Delta } f + 3 i m \right] \e_{2\pm} + \s^m \partial_m (4 \Delta +B) \e_{2 \mp} - \tfrac{i}{2}
e^{-2 \Delta} g^* \e_{1 \mp}\,.
\eea
The field equations of motion for our ansatz become
\bea
\label{3deom}
D (e^{4 \Delta -2 B} * \mathcal{A}) &=& e^{4 \Delta -2 B} \mathcal{P} \wedge * \mathcal{A}^* - i f g vol(M_3)\,, \nn
D ( e^{4 \Delta + 2 B} g) &=& e^{4 \Delta +2 B} g^* \mathcal{P} - i f \mathcal{A}\,, \nn
D( e^{8 \Delta +2 B} * \mathcal{P}) &=& - \tfrac{1}{4} e^{4 \Delta} \left[ e^{-2 B} \mathcal{A} \wedge * \mathcal{A} + e^{2 B} g^2 vol(M_3) \right]\,,
\eea
while the Bianchi identities may be expressed as
\bea
\label{bianchi}
d \mathcal{P} &=& 2 i Q \wedge \mathcal{P}\,, \nn
d \mathcal{A} &=& i Q \wedge \mathcal{A} - \mathcal{P} \wedge \mathcal{A}^*\,.
\eea
Observe that $g$ drops out from the Bianchi identitites and only its derivative
enters the equations of motion.

Finally, in orthonormal frame the Einstein equations may be written
\bea
&&\eta_{\mu \nu}\left[ - 4 m^2 - \nabla_m \nabla^m \Delta - 2 (4 \partial_m \Delta + \partial_m B) \partial^m\Delta \right] \nn
\label{einstein1} &&= \eta_{\mu \nu} \left[ - \tfrac{1}{4} f^2 e^{-8 \Delta} - \tfrac{1}{8} e^{-4 \Delta} (|g|^2 + e^{-4 B} \mathcal{A}_m \mathcal{A}^{*m})\right] \\
&& \delta_{\a \b} \left[ e^{-2 B} - \nabla_m \nabla^m (\Delta + B) - 2 (4 \partial_m \Delta + \partial_m B)(\partial^m \Delta + \partial^m B ) \right],  \nn
\label{einstein2} &&= \delta_{\a \b} \left[  \tfrac{1}{4} f^2 e^{-8 \Delta} + \tfrac{3}{8} e^{-4 \Delta -4 B} \mathcal{A}_m \mathcal{A}^{*m}  - \tfrac{1}{8} e^{-4 \Delta}
|g|^2\right] \\
\label{einstein3} &&R_{mn} - 2 \nabla_n \nabla_m (4 \Delta+ B)  + 8 \partial_n \Delta \partial_m \Delta - 2 \partial_n B \partial_m B  - \delta_{mn} \left[ \nabla^p
\nabla_p \Delta + 2 \partial_p (4 \Delta + B) \partial^p \Delta \right],  \nn
&&= \mathcal{P}_m \mathcal{P}^*_n + \mathcal{P}_n \mathcal{P}^*_m + \tfrac{1}{4} f^2 e^{-8 \Delta} \delta_{mn}  + \tfrac{3}{8} |g|^2 e^{-4 \Delta} \delta_{mn} \nn
&&+ \tfrac{1}{8} e^{-4 \Delta - 4 B} \left[2 \mathcal{A}_m \mathcal{A}_n^* + 2 \mathcal{A}_n \mathcal{A}_m^* - \delta_{mn} \mathcal{A}_p \mathcal{A}^{*p}\right].
\eea
Here we have used $\mu, \nu$ for $AdS_5$, $\alpha, \beta$ for $S^2$ and finally $m,n$ for the remaining directions. We have taken the sphere's radius to be unity
while the radius of $AdS_5$ is $m^{-1}$.

Having made the conditions on any supersymmetric three-dimensional geometry explicit in this section, we turn our attention in the next section to the example of
Sasaki-Einstein where there are only two independent spinors, and not four, as in the general case.

\section{$M_5$ Sasaki-Einstein}
\label{sec4}
In this section, by way of a warm-up, we address what happens when we set one of the $D=5$ spinors $\xi_i$ to zero and the $M_5$
geometry satisfies the Sasaki-Einstein Killing spinor equation. We will see from the spinor bilinear analysis below that one encounters multiple Killing
directions, where the warp factor $B$ depends on some of the Killing directions. From the Killing spinor equation on $M_3$ we show that $M_3$ is isomorphic
to $S^3$ and proceed to construct explicitly the $S^3$ Killing vectors from the vector bilinears. Back in five-dimensions, this translates into $M_5$ being
simply $S^5$ or some quotient (for example \cite{Aharony:1998xz}).

%Promoting in the spirit of \cite{GMSWm6,LLM,james}, rules out the existence of a geometry.

%\subsection{First impressions}
We begin by recalling the observation in~\cite{james} that the Sasaki-Einstein geometries correspond to choosing $\xi_2 = 0$ in $D=5$, which, in turn, sets
$\e_{2 \pm} = 0$ in our $D=3$ notation. In this setting, we can also use arguments presented in~\cite{james} showing that the three-form flux, $G$, is zero,
and the axion and dilaton are simply constants when $M_5$ is compact\footnote{{Harmonic functions $f$ satisfy the maximum principle: if $K$ is any compact subset
of a connected set $U$, then $f$, restricted to $K$, is a constant.}}. The only remaining non-constant scalar is then the warp factor, $B$, with the overall
ten-dimensional warp factor, $\Delta$ becoming a constant
\be
\label{Deltawarp}
e^{-4 \Delta} f = 4m.
\ee
As an aside, observe that, as $d \Delta = 0$,  (\ref{Deltawarp}) is consistent with the Einstein equation in the $AdS_5$ directions (\ref{einstein1}).

With these simplifications, the three-dimensional supersymmetry conditions reduce to
\bea
\label{d3kse1}
&&\nabla_{m} \e_{\pm} + \tfrac{i}{2} m \s_m \e_{\mp} = 0, \\ \label{d3kse2}
&&\left[ \pm i e^{-B} + i m  \right] \e_{\pm} + \s^m \partial_m B \e_{ \mp} = 0,
\eea
where we have dropped redundant subscripts and replaced $D$ with $\nabla$ as now $\mathcal{Q} = 0$. The familiar reader will identify above the Killing
spinor equation on $S^3$ \cite{s3kse}, a fact that we will return to soon.

In the present case, the relevant scalar spinor bilinears are $S_1, S_2, T_1$ and
$U_1$, while the vector bilinears are $K^1, K^2, L^1, M^1, M^2 $ and $N^1$ ({\it cf. } equations~(\ref{scalbilinear})
and~(\ref{vecbilinear}) in the appendix). Using equation~(\ref{d3kse1}) it is possible to determine the scalar bilinear differential  conditions
\bea
\label{se5stor1}
d S_1 &=& 0\,, \\
\label{se5stor2}d S_2 &=& m \Im(L_1)\,, \\
\label{se5stor3}d T_1 &=& - i m K^2\,, \\
\label{se5stor4}d U_1 &=& - i m M^2\,.
\eea
We can also determine the following algebraic expressions from (\ref{d3kse2})
\bea
\label{se5const1}
S_2 &=& - m e^{B} S_1\,, \\
\label{se5const2} \Re(T_1) &=&  0\,.
\eea
Above $\Re$ and $\Im$ denote the real and imaginary parts of a complex number.	 
Using the algebraic and differential KSEs one can show that show that $K^1, ~ \Re(L^1), ~ M^1$ and $N^1$ are Killing directions. For the moment, we postpone any attempt to determine the relationship between these Killing directions. In general, we have 6 Killing vectors as $M^1$ and $N^1$ are complex, however we stress that some of these Killing vectors may be trivially zero for a particular choice
of the spinors, while it is also possible that they align as noted in~\cite{eoin1}.
To have a geometry dual to ${\cal N}=2$ SCFTs, we require the existence of a single
overall $U(1)$ isometry direction from the above list.

These directions will generate symmetries of the overall spacetime, presenting us with candidate $U(1)$
R-symmetry directions, provided the warp factor, $B$, is
independent of the Killing directions. From (\ref{d3kse2}), we may determine the following relationships
\bea
\label{Bdep}
i_{K^1} d B &=& - e^{-B} \Im(T_1)\,, \nn
i_{\Re(L^1)} d B &=& 0\,, \nn
i_{M^1} d B &=& i e^{-B} U_1\,, \nn
i_{N^1} d B &=& 0\,.
\eea
If we demand that these directions correspond to  R-symmetries (global $U(1)$'s),
%are isometries of the full solution%
we may show that there is no geometry. To see this, note that from (\ref{se5const2}) and (\ref{Bdep}),
we have $T_1 = 0$ if $K^1$ is either zero or an R-symmetry direction. The derivative of $T_1$,
\be
d T_1 = - i m K^2\,,
\ee
then tells us that $K^2$ is zero. Finally, we can see that $K^1$ is also zero from (\ref{K1zero}). Together $K^1 = K^2 = 0$ tell us that $\e_{+} = \e_{-} = 0$, thus ruling
out a solution. A similar conclusion may be reached working with $U_1$, $M^1$ and $M^2$.

In fact, one may show  directly that the $D=3$ KSEs imply that
the space $M_3$ is a three-sphere. Evoking the integrability relationship
\be
\nabla_{[m} \nabla_{n]} \e_{\pm} = \frac{1}{8} R_{mnpq} \sigma^{pq} \e_{\pm},
\ee
one can show that
\be
m^2 \epsilon_{mnp} \sigma^p \e_{\pm} = \frac{1}{2} R_{mnq_1 q_2} \epsilon^{q_1 q_2}_{~~~q_3} \sigma^{q_3} \e_{\pm}.
\ee
This implies that $M_3$ is a manifold of constant curvature. Then, it is well known \cite{Kobayashi}, that for any $n$-dimensional connected, complete Riemannian manifold
$M$ of constant curvature $\tfrac{1}{a^2}$, the universal covering manifold of $M$ is isomorphic to a sphere of radius $a$, $M = S^n/G$, where $G$ denotes some finite subgroup of $O(n+1)$ that acts freely.

Having confirmed that there is one smooth (maximally) supersymmetric geometry that satisfies the above conditions on the spinor bilinears, it is an instructive exercise
to recover $M_3 = S^3$ from the conditions on the spinor bilinears directly. We start by dropping the requirement $\Im(T_1) = 0$, thus allowing the warp factor $B$ to depend
on the Killing directions on $M_3$, while for simplicity also setting $m=1$. Observe that the 6 Killing vectors noted earlier now may be interpreted as the generators of
$SO(4)$, the symmetry of the three-sphere. We will now show that orthonormal frame $e^i$, which may be read off from the vector bilinears $\bar{\e} \sigma_i \e  e^i$, is
simply that corresponding to a three-sphere.

We begin by rotating $K^1$ and $K^2$ into the $e^1, e^2$ plane. Since these are just two vectors, it is always valid to do this. After rotation, the conditions
$\bar{\e}_{\pm} \s_3 \e_{\pm} = 0 $ may be satisfied by writing
\be
\e_+ = r_1 \left( \begin{array}{c} e^{i \phi_1} \\ e^{i \phi_2} \end{array} \right), \quad \e_- = r_2 \left( \begin{array}{c} e^{i \psi_1} \\ e^{i \psi_2} \end{array} \right),
\ee
where $r_i$ correspond to the norms of the spinors. The norms on the spinors may be determined from (\ref{se5const1}) and constant $S_1 = 1$:
\be
r_1 = \sin \frac{\theta}{2}, \quad r_2 = \cos \frac{\theta}{2}.
\ee
It is worth noting that with this choice $e^{B} = \cos \theta$, so that the overall $S^5$ is the following fibration
\be
ds^2(S^5) = \cos^2 \theta ds^2(S^2) + d \theta^2  + \sin^2 \theta ds^2(S^2).
\ee

As $B$ is just a function of $\theta$, we align the $\theta$-direction with $e^1$. By combining (\ref{Bdep}) and the knowledge that $\Re(L_1)$ and $N_1$ are not
along $e^1$, $K^{1} \cdot \Re(L^1) = 0 $ from the Fierz identity, and $d (e^B) = - \Im(L^1)$ from (\ref{se5stor2}) and (\ref{se5const1}), one finds consistent
conditions that whittle down the angles to two. The form of the spinors is then
\be
\e_+ = \sin \frac{\theta}{2} \left( \begin{array}{c} e^{i \phi_1} \\ e^{i \phi_2} \end{array} \right), \quad \e_- = - i  \cos \frac{\theta}{2} \left( \begin{array}{c}
e^{i \phi_2} \\ e^{i \phi_1} \end{array} \right).
\ee

The $e^i$ labeling the orthonormal frame may then be read off from $d (e^B) = - \Im(L^1)$, (\ref{se5stor3}) and (\ref{se5stor4}), leading to the following
\be
e^1 = - d \theta, \quad e^2 = \sin \theta (d \phi_1 - d \phi_2), \quad e^3 = \sin \theta \sin(\phi_1 - \phi_2) (d \phi_1 + d \phi_2).
\ee
A redefinition
\bea
\phi = \phi_1 - \phi_2, \quad \psi = \phi_1 + \phi_2,
\eea
recovers the metric on $S^3$ with unit radius. As another consistency check we work out the Killing vectors for the space:
\bea
 K^1 &\rightarrow&  - \cos \phi \partial_{\theta} + \sin \phi \cot \theta \partial_{\phi} \nn
 \Re(L^1) &\rightarrow& - \partial_{\psi} \nn
 M^1 &\rightarrow& - \sin \phi \cos \psi \partial_{\theta} - \cot \theta (\cos \phi \cos \psi \partial_{\phi} -\frac{\sin \psi}{\sin \phi} \partial_{\psi} ),  \nn
&\rightarrow& - \sin \phi \sin \psi \partial_{\theta} - \cot \theta (\cos \phi \sin \psi \partial_{\phi} +\frac{\cos \psi}{\sin \phi} \partial_{\psi} ),  \nn
 N^1 &\rightarrow&   \sin \psi \partial_{\phi} + \cos \psi \cot \phi \partial_{\psi}, \quad -\cos \psi \partial_{\phi} + \sin \psi \cot \phi \partial_{\psi}.
\eea
These are simply the generators of $SO(4)$. Note neither $\Re(L^1)$ nor $N^1$ have components along the $\theta$ direction, a fact that is consistent with
(\ref{Bdep}).  Observe also that $K^1$ and $\Re(L^1)$ are commuting. We will see this is the case later when we turn to the more general case.

Finally, before moving onto the general case, we remark that multiple Killing directions also appeared in \cite{ads3s2}, and, in particular, the Lie derivative
of a similar $S^2$ warp-factor with respect to one Killing direction was found to be non-zero. Though this Killing direction was consistently removed from the
supersymmetry conditions and subsequent analysis led to a class of known geometries \cite{GOMW, Kim3}, in the light of observations here, it would be interesting
to revisit that example and retain all the Killing vectors.

\section{A tale of two Killing vectors}
\label{sec5}

In this section we consider the general form of our ansatz and start by identifying the candidate
Killing directions.  Recall that a $U(1)$ Killing direction (denoted $K_5^{GMSW}$) exists in
$D=5$~\cite{james}
for a general $M_5$ and generic values of all fields. As a result, we naturally expect this $U(1)$
direction to descend from $D=5$ to give a solution to the Killing equation in $D=3$. In fact, it is
possible to identify the combination $X = K^1 + K^3$ and the combination $Y = \Re(L^1+ L^6)$ (see appendix for the
explicit definitions of the spinor bilinears) as solutions to the Killing equation for all values of the
spacetime fields. The $D=5$ Killing direction $K_5^{GMSW} = \tfrac{1}{2}(\bar{\xi}_1 \gamma^m
\xi_1+\bar{\xi}_2
\gamma^m \xi_2)$ can be decomposed using our ansatz our ansatz~(\ref{spinor5to3}) to give
\be
\label{k5asxy}
K_5^{GMSW} = (\bar{\chi}_+ \sigma_3 \chi_+) X + (\bar{\chi}_+ \chi_+) Y.
\ee
In other words, $K_5^{GMSW}$ is a particular linear combination of the two Killing directions we found
in $D=3$; note that the coefficients in the linear combinatoins are independent of $M_3$.

Our analysis so far shows that $X$ and $Y$ are Killing directions on $M_3$. Since we only expect one $U(1)$ R-symmetry, one possible interpretation might be that
either $X$ \textit{or} $Y$ generates the R-symmetry, with the other vector simply corresponding to a $U(1)$ isometry direction unrelated to the R-symmetry. Such an
example exists in the uplift of the $Y^{p,q}$ spaces~\cite{ypq} to M-theory~\cite{GMSWm6} where the Reeb vector associated with the R-symmetry combines with another
$U(1)$ from $Y^{p,q}$ to give the R-symmetry direction in M-theory. See appendix C of~\cite{eoin3} for further discussion on this subject.

%It is likely that both $X$ and $Y$ are Killing and overall $U(1)$ isometry directions corresponding to the R-symmetry, %however one has to allow for the case where two $U(1)$
%Killing directions combine to give the R-symmetry, so below we consider the case when either $X$ or $Y$ are R-symmetry %generators. This phenomenon where two Killing directions
%combine to give the R-symmetry is quite notable in the uplift of the $Y^{p,q}$ spaces \cite{ypq} to M-theory \cite{GMSWm6} %where the Reeb vector associated with the R-symmetry
%combines with another $U(1)$ from $Y^{p,q}$ to give the R-symmetry direction in M-theory. See appendix C of \cite{eoin3} for %further discussion on this subject.
%\vspace{2mm}

Let us first check then, under what conditions $X$ may be promoted to a $U(1)$ isometry of the overall solution. $X$ corresponding to an overall $U(1)$ is an important
prerequisite for it being identified as an R-symmetry direction. We proceed to calculate the Lie derivative of the various fields and warp factors with respect to $X$.
After some arithmetic we find
\bea
\label{Kkilling2}
\mathcal{L}_{X} \Delta &=& 0\,, \nn
\mathcal{L}_{X} B &=& - 2e^{-B} \Im(T_1 + T_6)\,, \nn
\mathcal{L}_{X} \mathcal{P} &=& 0, ~~ \Rightarrow \mathcal{L}_{X} \mathcal{Q} = 0\,,\nn
\mathcal{L}_{X} \mathcal{A} &=& 0\,,\nn
i_{X} d  g &=& 0\,.
\eea
To derive these equations we use the algebraic conditions~(\ref{alg1})~-~(\ref{alg8}),
equation~(\ref{realt1t6}), equation of motion for $g$~(\ref{3deom}), the Bianchi
identity~(\ref{bianchi}) as well as $i_X \mathcal{P}=0$.
%The Lie derivative of $\mathcal{A}$ with respect to $X$ is a little trickier to work out, so we begin by recording its %contraction with $X$ from (\ref{alg7}) and (\ref{alg8}):
%\be
%\label{AX}
%i_{X} \mathcal{A} = g e^{2 B} \Re(T_1 + T_6)\,.
%\ee
%As the RHS vanishes as a result of (\ref{realt1t6}), one just has to guarantee that $i_{X} d \mathcal{A} = 0 $, and this may %be seen to hold from the Bianchi (\ref{bianchi}),
%therefore ensuring that $\mathcal{L}_{X} \mathcal{A} = 0$.
%
As a result we conclude that the Killing direction, $X$, can be promoted to a symmetry of the full solution provided
\be
\label{imt1t6}
\Im(T_1 + T_6) = 0\,.
\ee
We now show that this condition implies either $f = 0 $ or $g = 0$ or $X=0$. At each step we
will assume that both $f$ and $g$ are non-zero and will ultimately show that this implies $X = 0$. In
section~\ref{sec6} below we will then consider the case $X\neq 0$ and $f=g=0$, while in
section~\ref{sec7} below we will consider the case $X=0$.
Firstly, equations~(\ref{imt1t6}), (\ref{Xrsym1}) and (\ref{Xrsym2}) imply $T_2 = T_5 = 0$. Combining
this result with equations~(\ref{Xrsym3}), (\ref{scaltor5}) and (\ref{scaltor6}) one finds that $L^3 =
L^4$ and $K^2=K^4$. Using the latter relation, together with equation~(\ref{imt1t6}), and $T_2 = T_5 =
0$ in equations~(\ref{K1zero}) and~(\ref{nosol2}) we find that $X=K^1+K^3=0$. To sum up we have shown
that $X$ is an isometry of the full solution only when $f$ or $g$ is zero.

%\noindent
%\textbf{Y R-symmetry}
We now turn to the analysis of $Y$. One can be show that
%\bea
%i_{Y} d \Delta &=& \tfrac{i}{8} e^{-2 \Delta} \Re[g(T_3 + T_4)] -
%\tfrac{i}{8} e^{-2 \Delta - 2 B} \Re[ \mathcal{A}^m
%(L_2+L_5)_m ] \nn &+& \tfrac{i}{4} e^{-4 \Delta } f
%(S_1 - S_3) - i m (S_1 + S_3), \\
%i_{Y} d B &=& - i e^{-B} (S_2 + S_4) + im (S_1 + S_3)
%- \tfrac{i}{2} e^{-4 \Delta} f (S_1 - S_3) \nn &+& \tfrac{i}{2} e^{-2
%\Delta -2 B} \Re [\mathcal{A}^m (L_2 + L_5)_m ].
%\eea
%Note the LHS of both these expressions are real, while the RHS are complex. As a result
\bea
\mathcal{L}_{Y} \Delta &=& 0 \,,\\
\mathcal{L}_{Y} B &=& 0\,,\\
\mathcal{L}_{Y} \mathcal{P} &=& 0\,,\\
\mathcal{L}_{Y} g &=&
\label{killconst1}
f g (S_1 + S_3) = 0\,, \\
\label{killconst2}
\mathcal{L}_{Y} \mathcal{A} &=&
2 e^{4 \Delta + 2 B} (g^* \mathcal{P} -2 d \Delta g) - i f \mathcal{A} = 0\,.
\eea
In deriving these identitites we have used equations~(\ref{diff1}) and~(\ref{diff2}), the algebraic conditions~(\ref{alg1})~-~(\ref{alg8}), the Bianchi identity~(\ref{bianchi}), equations of
motion (\ref{3deom}), equations~(\ref{scaltor1}) and (\ref{scaltor2}), as well as reality properties of the spinor bilinears.~\footnote{For example, the fact that $S_1\,,\dots\,,S_4$ are real.}
%In a similar fashion to above, it
%Furthermore, eliminating $\mathcal{A}$ terms from the RHS,
%we find a condition consistent with (\ref{consist1}) and
%(\ref{consist2}).
%It is also easy to show the following
%\bea
%i_{Y} \mathcal{P} &=&  i_{Y} \mathcal{Q} = 0, \nn
%i_{Y} \mathcal{A} &=& 2 g e^{2 B} (S_1 + S_3).
%\eea
%$\mathcal{L}_{Y} \mathcal{P} = 0$ then follows from the
%Bianchi (\ref{bianchi}) without any concern. The lower line is more constraining
%and it is possible to use the fact that
%$S_1+S_3$ is a constant from (\ref{scaltor1}) and (\ref{scaltor2}), in addition to the equations of
%motion (\ref{3deom}) to show that for $Y$ to be symmetry of the overall solution, we require
%\bea
%\mathcal{L}_{Y} g &=& \label{killconst1} f g (S_1 + S_3) = 0, \\
%\label{killconst2} \mathcal{L}_{Y} \mathcal{A} &=& 2 e^{4 \Delta
%+ 2 B} (g^* \mathcal{P} -2 d \Delta g) - i f \mathcal{A} = 0.
%\eea

It is worth noting that (\ref{killconst1}) is a particularly strong constraint. $S_1+S_3$ is a constant related to the norm of the spinors and
cannot be zero, so either $f=0$ or $g = 0$~\footnote{Note we have derived (\ref{killconst1}) on the assumption that $d g \neq 0$ and this constraint
on $f$ and $g$ disappears once $g$ is a constant. However, when $g$ is constant it is possible to derive a new version of (\ref{killconst2}) stating
\be \mathcal{L}_{Y} \mathcal{A} = 0 \Rightarrow 2 d B g - i \mathcal{Q} g + \mathcal{P} g^* = 0. \ee Observe now that the warp factor $B$ is charged
under $X$ whereas $\mathcal{P}$ and $\mathcal{Q}$ are not. This expression can only be consistent if either $g=0$, or $\mathcal{L}_X B = 0$, which we
have seen from the last section also implies either $f=0$ or $g=0$. So, the implication of one of $f$ or $g$ being zero does not change when $g$ is a constant.}.

Our two cases have now coalesced and we proceed to show that \textit{both} $f$ and $g$ are zero. Now if one adopts $f=0$, one sees from (\ref{f01}) and (\ref{f02})
that $S_2 = S_4$, while (\ref{f03}) and (\ref{f04}) in turn confirm that $g$ is also zero. Therefore, $f=0 \Rightarrow g=0$.

From the relations derived in the appendix, it is not obvious that $g=0$ implies $f=0$. However, as is clear from the equations of motion (\ref{3deom}), when $g=0$,
either $f =0$ or $\mathcal{A} = 0$. When $\mathcal{A} = 0$, the type IIB three-form flux vanishes and it is possible to show that both the axion and dilaton are harmonic.
Then, if $M_3$ is compact, as in the case of most interest to AdS/CFT, one deduces that both are constants {using the maximum principle of harmonic forms. }

Returning to five-dimensions, we are now on the cusp of declaring that the five-dimensional geometry is Sasaki-Einstein when $\mathcal{A} = g = 0$ i.e. no three-form flux.
However, before making such a statement, we require one of the spinors $\xi_i$ appearing in (\ref{sone}) to (\ref{ssix}) to vanish. Recall that it was shown in \cite{james}
that $\xi_2 = 0$ implies the five-dimensional space is Sasaki-Einstein. Indeed, when the three-form flux is zero, it is a simple exercise to show that (\ref{sthree}) and
 (\ref{sfour}) together imply $d \Delta (\bar{\xi}_1 \xi_1 + \bar{\xi}_2 \xi_2) = 0$. In other words $ \Delta$ is a constant and the Killing spinor equations can only be
satisfied if one of the $\xi_i$ is zero. The property that the five-dimensional space is then Sasaki-Einstein follows and the analysis reverts to that in the previous section.

In summary, in this section we have shown that requiring either of the Killing directions $X$ or $Y$ to correspond to a global $U(1)$ symmetry corresponding to the
R-symmetry,
implies that $f=g = 0$. We turn our focus to the analysis of these spacetimes in the next section.

\section{Geometries with $f=g = 0$}
\label{sec6}

In the last section we showed that the Killing vectors $X$ and $Y$ on $M_3$ could be promoted to global $U(1)$ isometries provided we restrict our ansatz by setting
$f=g = 0$.  In this section we restrict our ansatz to $f=g=0$.~\footnote{All equations in this section are derived with $f=g=0$.} We show that, for such backgrounds,
requiring $X$ or $Y$ to be a global $U(1)$ symmetry results in $X$ and $Y$ being equal to zero. The same result holds also if one entertains the idea that a linear combination of $X$ and $Y$ correspond to the R-symmetry direction. In the next section we will show that the condition $X=Y=0$ together
with
the KSEs in fact sets the Killing spinors to zero, leaving us with no solutions.

We start, by further examining
the Killing directions when $f=g=0$. As noted in section~\ref{sec4} above, when some of the field content is removed, extra Killing directions may emerge. After
re-examining the KSEs with $f=g=0$, we find that in the present case there are no new Killing directions.
We now examine the relationship between $X$ and $Y$ by checking to see if the Killing vectors commute. This may be done by calculating the Lie derivative $\mathcal{L}_{X} Y$. Making use of the Fierz identity and the identities $\s_{mp} \s_q \s^m = -\s^m \s_q \s_{mp} = 2 \delta_{pq}$, one can show indeed that $i_X d Y = i_Y d X = 0$ and that the vectors commute
\be
\mathcal{L}_{X} Y = \left[X\,,Y \right]=0\,.
\ee
Recall that in the LLM geometry~\cite{LLM}, which has a similar ansatz,  two Killing directions align to give the R-symmetry~\cite{eoin1}. In the IIB geometry we are presently considering, it
is possible to use the Fierz identity to show that $X$ and $Y$ are in fact
orthogonal~\footnote{Firstly, one shows that $K^1 \cdot \Re(L^1)
= K^3 \cdot \Re(L^6) = 0$ using~(\ref{realt1t6b}) below, while the cross-terms can be confirmed to vanish by using two iterations of the Fierz identity,~(\ref{realt1t6b}) again and~(\ref{t3t4}) below.}
\be
X\cdot Y =0\,.
\ee

The algebraic constraints~(\ref{alg1})~-~(\ref{alg6}),
together with $f=g=0$ can be used to obtain the following spinor bilnear relations
\bea
\label{norm1} S_1 &=& S_3, \quad S_2 = S_4, \\
\label{t3t4} T_3 + T_4 &=& 0, \\
\label{u3u4} U_3 + U_4 &=& 0, \\  \label{realt1t6b} \Re(T_1) &=& \Re(T_6) = 0, \\
\label{u2u5} U_2 - U_5 &=& - 3 m e^{B} (U_2 +U_5), \\
\label{nosol1} T_2 - T_5 &=& - 3 m e^B (T_2 + T_5), \\
\label{norm2} S_2 &=& - 3 m e^B S_1.
\eea
%Once again we adopt the choice $S_1 + S_3 = 1$ and proceed to determine the differential conditions satisfied by the %bilinears. As these are quite lengthy, we
%collect them in appendix C and here confine our attention to some comments.
%
%Firstly, some of the torsion conditions appearing in appendix C
%have not been rewritten using the algebraic constraints so as %to avoid introducing Hodge duals
%which appear frequently as we are working in $d=3$. In addition to the
%constraints in (\ref{norm1}) - (\ref{norm2}), one can %further infer
%\bea
%D (T_2 + T_5) &=& 0, \\
%\label{diffu2u5} d (U_2 + U _5) &=& 0, \\
%\Im(L^1) &=& \Im(L^6),
%\eea
%from the scalar bilinear differential conditions.

From the vector bilinear differential conditions, we note that the two Killing vectors $X$ and $Y$ satisfy
\be
\label{xy}
d X = 2m * Y, \quad d Y = 2 m *X.
\ee
As an aside, observe that $d * X = d * Y = 0$, so that the Laplacian acting on $X$ or $Y$ becomes $*d * d X = 4m^2 X$. Interestingly, if $M_3$ is a three-dimensional
compact Riemannian Einstein space normalised such that $R_{mn} = 2 m^2 g_{mn}$, the Laplacian eigenvalues $\kappa$ satisfy $\kappa \geq 4 m^2$ with saturation happening
when the one-form is Killing \cite{dewolfe}. This precisely what one notes above for $X$ and $Y$.

In addition, we also see from the vector torsion conditions that $X$ satisfies
\be
\label{dx}
d X = 2 d B \wedge X.
\ee
%This tells us that $d B$ cannot align with $X$, otherwise $X = Y = 0$ from (\ref{xy}).
%In addition $\Im(L^1) = \Im(L^6)$ satisfies
%\be
%d ( e^{4 \Delta} \Im(L^1))  =  0.
%\ee
%This is consistent with (\ref{DeltaB}).

So what have we learned about $M_3$? We have seen that $M_3$ has two orthogonal Killing directions, so we can think of $M_3$ having three directions: $X$, $Y$ and an
additional direction parameterised by, say, $\theta$. Observe that $X$ and $Y$ are coupled, so we cannot set one of them to zero without also setting the other to
zero. Presently, we explore the possibility that $X\neq 0\neq Y$ and that either $X$ or $Y$ corresponds to the $U(1)$ R-symmetry (and so a global $U(1)$ symmetry)
while the other is simply a generic
isometry direction.

\vspace{2mm}
\noindent
\textbf{$X$ is a global $U(1)$ symmetry direction} \\
Recall from section~\ref{sec5} that $X$ can be a
global $U(1)$ symmetry if $\Im(T_1 + T_6) = 0$. As $T_1$ and $T_6$ are both pure imaginary, combining (\ref{nosol1}), (\ref{t1diff}), (\ref{t6diff}), (\ref{K1zero})
and (\ref{nosol2}), together with $f=g=0$ one can show that
\be
X = K^2 + K^4 = \mathcal{A} T_2 = \mathcal{A} T_5 = 0\,.
\ee
But equation~(\ref{xy}) then implies that $Y=0$. So requiring that $X$ is a global $U(1)$ symmetry implies that $X=Y=0$.

\vspace{2mm}
\noindent
\textbf{$Y$ is a global $U(1)$ symmetry direction} \\
Instead let us see what happens when we require that $Y$ be a global $U(1)$ symmetry. Recall that presently $M_3$ is parametrised by three directions $X$, $Y$ and
$\theta$. Equations~(\ref{Kkilling2}) and~(\ref{dx}) imply that $B$ depends on $X$ and $\theta$, but is independent of the $Y$-direction.
%Observe also that (\ref{DeltaB}),
%(\ref{Xrsym3}), (\ref{closed1}) and (\ref{closed2}) were all derived
%independently but are internally consistent in the sense %that one can use (\ref{norm1}) and
%(\ref{norm2}) to show that (\ref{DeltaB}) and (\ref{Xrsym3}) are the same,
%while at the same time, the expressions may be %added.
We also recall, from the results in section~\ref{sec5} that both $\mathcal{A}$ and $d \Delta$ have only
components along the $\theta$-direction. Then from equation~(\ref{closed2}) one can see that
$\Im(L^1+L^6)$
has no component along the $X$ or $Y$ directions. This last fact, in conjunction with a similar
observation for $d\Delta$, implies, via equation~(\ref{DeltaB}), that $B$ has no $X$-dependence
afterall. As a result
\be
\Im(T_1+T_6)=0\,,
\ee
and so $X=Y=0$.~\footnote{To get around this argument, one may attempt to set $\Im(L^1+L^6) = 0$. However, this
is catastrophic as $d (4 \Delta +B) = 0$ with $f=g=0$ means from equations~(\ref{alg5}) and~(\ref{alg6})
that $B$ is also a constant, $e^{-B} = \pm 3m $ and that two spinors are zero.}

\vspace{1cm}
To summarise, in this section we have shown that one cannot require $X$ or $Y$ or any linear combination to be a global $U(1)$ symmetry direction. In the next section we will investigate the
possibility that both $X$ and $Y$ are zero.

\section{New R-symmetry directions?}
\label{sec7}

In the last section we saw that the KSEs do not allow for $X$ or $Y$ to be global $U(1)$ symmetry directions, and, hence, they cannot be $U(1)$ R-symmetry directions
either.
We are left with the posibility that $X=Y=0$. It is easy to show that these conditions alone do not
force all the Killing spinors to be zero. Rather, they merely reduce the number of independent Killing
spinor components. As such, {\it a priori} it appears possible to consider the KSEs with this restricted
choice of Killing spinors, and to re-start a search for solutions. Such a possibility appears already in
the ${\cal N}=1$ setting discussed in~\cite{james}. There a unique Killing vector, $K^{GMSW}_5$, was
found;~\footnote{Throughout this section we shall use the superscript GMSW to denote notation
from~\cite{james}.} and
one may wonder whether solutions to the KSEs exist for which $K^{GMSW}_5=0$, and instead a new $U(1)$
R-symmetry Killing vector emerges in this restricted setting.~\footnote{We are grateful to Dario
Martelli and James Sparks for a number of illuminating discussions during the Benasque Strings 2011
workshop on this subject.} In this section we will show that this apparent loophole in fact does not
lead to any new solutions in the ${\cal N}=1$ case. In particular we will show that setting
$K^{GMSW}_5=0$,
together with the KSEs imply that the $D=5$ Killing spinors $\xi_i$ have to be identically equal to zero.
Since the ${\cal N}=2$ case we are considering here is a special case of the ${\cal N}=1$ geometries
studied in~\cite{james}, and since $X=Y=0$ implies that $K^{GMSW}_5=0$, the argument presented in this
section
will also imply that no solutions exist when $X$ and $Y$ are set to zero.

From~\cite{james} we see that the norm of $K^{GMSW}_5$ is
\be
|K^{GMSW}_5|^2 = \sin^2 \zeta^{GMSW} + |S^{GMSW}|^2\,,
\ee
thus implying that both $S^{GMSW}$ and $\sin \zeta^{GMSW}$ are zero. Then from
equation~(3.15) of the same paper we have that the complex vector $K^{GMSW}$ is zero.
Furthermore, using equation~(3.22) of~\cite{james}, it is possible to show that the
two-form bilinear $U^{GMSW}$ is also zero. These extra constraints on the spinors may be summarised as
\bea
\label{K50cond}
\sin \zeta^{GMSW} &=& \tfrac{1}{2}(\bar{\xi}_1 \xi_1 -
\bar{\xi}_2 \xi_2) = 0\,, \nn
S^{GMSW} &=& \bar{\xi}_2^c \xi_1 = 0\,, \nn
K_5^m{}^{GMSW} &=& \tfrac{1}{2} (\bar{\xi}_1 \g^m \xi_1 + \bar{\xi}_2
\g^m \xi_2) = 0\,, \nn
K_m^{GMSW} &=&
\bar{\xi}_1^c \g_m \xi_2 = 0\,, \nn
iU_{mn}^{GMSW} &=& \tfrac{1}{2} (\bar{\xi}_1 \g^{mn} \xi_1 + \bar{\xi}_2
\g^{mn} \xi_2) = 0\,.
\eea
We now show that these conditions imply $\xi_1 = \xi_2 = 0$. With the explicit gamma matrix
representation given in~\cite{james} we have
\be
C_5 = \mathbf{1} \otimes i \sigma^2\,.
\ee
One can now decompose $\xi_1$ and $\xi_2$ as
\be
\xi_1 = \left( \begin{array}{c} r \\ s \\ t \\ u \end{array}
\right), \quad \xi_2 = \left( \begin{array}{c} w \\ x \\ y \\z \end{array} \right),
\ee
where the components are in general complex. Now, by combining $S^{GMSW} = K^{GMSW} = 0$, one
establishes the following relations
\bea
tz &=& u y, ~~rx = sw, ~~tx = sy, \nn uw &=& rz, ~~ux = sz,
~~tw = ry\,.
\eea
Using the
remaining relationships it is easy to find that one component of $\xi_1$ or $\xi_2$ is zero. For
example, using $(K^3_5)^{GMSW} = (K_5^4)^{GMSW} = U_{13}^{GMSW} = U_{14}^{GMSW} = 0$, it is possible to
infer that
\be
z (|r|^2 + |w|^2)= 0\,,
\ee
so that either $z = 0$ or $r = w = 0$ is zero. Once one of the components in
$\xi_1$ or $\xi_2$ can be shown to be zero, the result that $\xi_1 = \xi_2 = 0$ is immediate from the
remaining relationships coming from $K_5^{GMSW} = U^{GMSW} = 0$. Therefore, we conclude that setting
$K^{GMSW}_5=0$, together with the KSEs, implies that the Killing spinors are zero.

In fact a similar argument can also be seen to hold for the ${\cal N}=2$ case that has been the focus of
this paper. When $X=Y=0$, we also have $K^{GMSW}_5=0$ ({\it cf.} equation~(\ref{k5asxy})). We can then
re-write
the conditions~(\ref{K50cond}), when reduced using the ansatz~(\ref{spinor5to3}). In particular, the
$U^{GMSW}_{mn}$ condition, with $m$ and $n$ in the directions of $S^2$ is particularly strong (using the
gamma matrix basis given in the appendix) and implies that the sum of the norms of $\epsilon_{i\pm}$ has
to be zero, and so the Killing spinors $\epsilon_{i\pm}$ have to be zero themselves.

This concludes our analysis. We have used the KSEs to search for Type IIB solutions with $AdS_5$ and
$S^2$ factors and R-symmetry $SU(2) \times U(1)$. We have found that no non-trivial solutions of such a type exist. Our result suggests
that if one can find Type IIB solutions which are holographic duals of four-dimensional ${\cal N}=2$
SCFTs, the $SU(2)$ R-symmetry will have to be realised in some non-geometric way which does not lead to
the presence of $SU(2)$ Killing vectors in the spacetime.

\section*{Acknowledgements} We would like to thank Jerome Gauntlett, Chris Hull, Oleg Lunin, Juan Maldacena, Hiroaki Nakajima, Dimitri Skliros,  Linda Uruchurtu, Oscar Varela, Hossein Yavartanoo,  and especially Dario Martelli and James Sparks for interesting discussions and sharing their insights with us. We are grateful to the organisers and participants of the Benasque Strings 2011 workshop, and the {\em Centro de Ciencias de Benasque Pedro Pascual} for providing a stimulating and productive atmosphere for the final stages of this project. E\'OC also expresses gratitude to the {\em Simons Center for Geometry and Physics} and the organisers of the Simons Summer Workshop on Geometry and Physics 2011 for generous hospitality while this draft received some final tweaks. The work of BS is supported by an EPSRC Advanced Research Fellowship.

\appendix
\section{Spinor and $\gamma$-matrix conventions}
We will use the spinor conventions of~\cite{james} (see also \cite{sohnius}) under the understanding that we are in addition
decomposing
the internal $M_5$ space into a direct product of $S^2$ and $M_3$. We begin by recalling that the $M_5$ gamma matrices $\g_i$ $i=1,\dots,5$ satisfy the following:
\bea
\label{C5cond}
\g_{i} &=& \g_i^{\dagger}, \nn
C_5^{-1} \g_i C_5 &=& \g_i^{T},
\eea
where $\tilde{D}_5 = C_5$ and $C_5^{*} = - C_5^{-1}$, $C_5 = - C_5^{T}$. In addition, $\g_{12345} = 1$. A five-dimensional spinor $\chi$, where $\chi^{c} = C_5 \chi^*$
satisfied $\chi^{cc} = - \chi$.

Then adopting the following choice for the decomposition of $\g_{i}$
\bea
\label{gammadecomp}
\g_{\a} &=& |\e_{\a \b}| \s_{\b} \otimes 1, \nn
\g_{m+2} &=& \s_3 \otimes \s_m,
\eea
where $\a =1,2$, $m =1,2,3$ and we have introduced somewhat awkward ordering so that $\g_{12345} = 1$ as in \cite{james}. One sees that a natural choice is
\be
C_5 = \sigma_1 \times \sigma_2.
\ee
Here we have taken $C_3 = \s_2$ so that $C_3^{-1} \s_m C_3 = - \s_m^{T}$ consistent with \cite{sohnius}. With this choice if $\chi$ is a Killing spinor on $S^2$, i.e. a solution to
\be
\nabla_{\a} \chi = i \frac{\s_{\alpha}}{2} \chi,
\ee
then it is easy to see that the conjugate $\chi^{c} = C_2 \chi^* = \s_1 \chi^*$ satisfies the same equation with the opposite sign.

\section{Bispinor relations}
Here we record the bilinears that appear in the analysis.
We label the scalars constructed out of our spinors in the following fashion,
\bea
\label{scalbilinear}
S_{1} &=& \tfrac{1}{2}(\bar{\e}_{1+} \e_{1+} + \bar{\e}_{1-} \e_{1-}),~~S_{2}= \tfrac{1}{2}(\bar{\e}_{1+} \e_{1+} - \bar{\e}_{1-} \e_{1-})\,, \nn
S_{3} &=& \tfrac{1}{2}(\bar{\e}_{2+} \e_{2+} + \bar{\e}_{2-} \e_{2-}),~~S_{4} = \tfrac{1}{2}(\bar{\e}_{2+} \e_{2+} - \bar{\e}_{2-} \e_{2-})\,, \nn
T_{1} &=& \bar{\e}_{1+} \e_{1-}, ~~T_2 = \bar{\e}_{1+} \e_{2+},~~T_3 = \bar{\e}_{1+} \e_{2-}\,, \nn
T_{4} &=& \bar{\e}_{1-} \e_{2+},~~T_5 = \bar{\e}_{1-} \e_{2-},~~T_6 = \bar{\e}_{2+} \e_{2-}\,, \nn
U_{1} &=& \bar{\e}_{1+}^c \e_{1-}, ~~U_2 = \bar{\e}_{1+}^c \e_{2+},~~U_3 = \bar{\e}_{1+}^c \e_{2-}\,, \nn
U_{4} &=& \bar{\e}_{1-}^c \e_{2+},~~U_5 = \bar{\e}_{1-}^c \e_{2-},~~U_6 = \bar{\e}_{2+}^c \e_{2-}\,,
\eea
and the vectors thus:
\bea
\label{vecbilinear}
K^1_{m} &=& \tfrac{1}{2} (\bar{\e}_{1+} \s_m \e_{1+} + \bar{\e}_{1-} \s_m \e_{1-}),~~K^{2}_m = \tfrac{1}{2}(\bar{\e}_{1+} \s_m \e_{1+} - \bar{\e}_{1-} \s_m \e_{1-})\,,
\nn
K^3_{m} &=& \tfrac{1}{2} (\bar{\e}_{2+} \s_m \e_{2+} + \bar{\e}_{2-} \s_m \e_{2-}),~~K^{4}_m = \tfrac{1}{2}(\bar{\e}_{2+} \s_m \e_{2+} - \bar{\e}_{2-} \s_m \e_{2-})\,,
\nn
L^{1}_m &=& \bar{\e}_{1+} \s_m \e_{1-}, ~~L^2_m = \bar{\e}_{1+} \s_m \e_{2+},~~L^3_m = \bar{\e}_{1+} \s_m \e_{2-}\,, \nn
L^4_m &=& \bar{\e}_{1-} \s_m \e_{2+},~~L^5_m = \bar{\e}_{1-} \s_m \e_{2-},~~L^6_m = \bar{\e}_{2+} \s_m \e_{2-}\,, \nn
M^1_{m} &=& \tfrac{1}{2} (\bar{\e}^c_{1+} \s_m \e_{1+} + \bar{\e}^c_{1-} \s_m \e_{1-}),~~M^{2}_m =
\tfrac{1}{2}(\bar{\e}^c_{1+} \s_m \e_{1+} - \bar{\e}^c_{1-} \s_m \e_{1-})\,, \nn
M^3_{m} &=& \tfrac{1}{2} (\bar{\e}^c_{2+} \s_m \e_{2+} + \bar{\e}^c_{2-} \s_m \e_{2-}),~~M^{4}_m = \tfrac{1}{2}(\bar{\e}^c_{2+}
\s_m \e_{2+} - \bar{\e}^c_{2-} \s_m \e_{2-})\,, \nn
N^{1}_m &=& \bar{\e}^c_{1+} \s_m \e_{1-}, ~~N^2_m = \bar{\e}^c_{1+} \s_m \e_{2+},~~N^3_m = \bar{\e}^c_{1+} \s_m \e_{2-}\,, \nn
N^4_m &=& \bar{\e}^c_{1-} \s_m \e_{2+},~~N^5_m = \bar{\e}^c_{1-} \s_m \e_{2-},~~N^6_m = \bar{\e}^c_{2+} \s_m \e_{2-}\,.
\eea
%\subsection{Useful relationships}
The relationships presented here may all be derived through either Fierz identities, or direct manipulation of the algebraic $D=3$ supersymmetry conditions (\ref{alg1}) -
(\ref{alg6}). We break them down into expressions involving scalars
\bea
\label{realt1t6} \Re(T_1) &=&  \Re(T_6) = 0, \\
\label{f01}  e^{-B} (S_2 - S_4) &=& m (S_1 -S_3) - \tfrac{1}{2} e^{-4 \Delta} f (S_1 + S_3), \\
\label{f02}
e^{-4 \Delta} f (S_1 + S_3) &=& 4 m (S_1 - S_3), \\
%\label{consist1}\Im [g (T_3 - T_4)] &=& 0, \\
%\Re(g T_2) &=&  \Re(g T_5) = 0, \\
%e^{-4 \Delta } f \Re(T_1 + T_6) &=& 4 m \Re(T_1 - T_6), \\
%e^{-B} (S_2 -S_4) &=& \tfrac{1}{2} e^{-4 \Delta} f (S_1+S_3) - 3 m (S_1 - S_3), \\
% e^{-B} (S_2 + S_4) &=& \tfrac{1}{2} e^{-4 \Delta} f (S_1 - S_3) -3 m (S_1 + S_3) \nn &+& \label{consist2} \tfrac{1}{2} e^{-2 \Delta} (g T_3 + g^* T_4^*), \\
\label{Xrsym1} e^{-2 \Delta} f (T_2 -T_5) &=& i g^* \Im(T_1+T_6), \\
\label{f03} 6m (T_3 + T_4) &=& e^{-2 \Delta} g^* (S_1 + S_3), \\
\label{f04} -2 e^{-B} (T_3 + T_4) + e^{-4 \Delta} f (T_3 -T_4) &=& e^{-2 \Delta} g^* (S_2-S_4), \\
%\Re(T_1) &=& \Re(T_6), \nn
%f \Re(T_1+T_6) &=& 0, \nn
\label{Xrsym2} 3 m (T_2 +T_5) &=& -  e^{-B} (T_2-T_5).
%- 2 e^{-B} \Re(T_1) &=& \tfrac{1}{2} e^{-2 \Delta} \Re(gT_2 - gT_5), \nn
%3 m \Re(T_1 + T_6) &=& \tfrac{1}{2} e^{-2 \Delta} \Re( g T_2 + g T_5), \nn
%g U_3 &=& f U_3 = 0, \\
%U_3 &=& -U_4, \\
%e^{-B} (U_5- U_2) &=& 3 m (U_2 + U_5), \\
%e^{-4 \Delta} f (U_2 -U_5) &=& e^{-2 \Delta}  (g U_6 + g^* U_1),
\eea
and expressions involving vectors
%From (\ref{alg5}) and (\ref{alg6}) we deduce the following:
\bea
\label{DeltaB} (S_1 + S_3) d (4 \Delta + B) &=& - e^{-B} \Im(L^1 + L^6), \\
%2 (S_2+S_4) d (4 \Delta +B) &=& 6m \Im(L^1+L^6) + e^{-4 \Delta} f \Im(L^6-L^1), \\
\label{Xrsym3} 2 (T_5-T_2) d (4 \Delta +B) &=& 6 im (L^3 - L^4) -i e^{-2 \Delta} g^* (K^2-K^4), \\
%\label{t2t5vec} 2 (T_5 + T_2) d (4 \Delta + B) &=& 2 i e^{-B} (L^3 - L^4) - i e^{-4 \Delta} f (L^3 + L^4) \nn &+& i e^{-2 \Delta} g^* (K^1 - K^3), \\
%2 (T_3 + T_4) d (4 \Delta +B) &=& i e^{-2 \Delta} g^* \Re(L^1-L^6)- i e^{-4 \Delta} f (L^2+L^5), \\
\label{K1zero} 2 (T_1 +T_6) d (4 \Delta +B) &=& - 6 im (K^2 +K^4) + i e^{-4 \Delta} f (K^2 -K^4)\nn &-&2 i e^{-B} (K^1 +K^3).
\eea
From (\ref{alg1}) and (\ref{alg2}) one finds
\bea
\label{nosol2} 2 d B \Im (T_1+T_6) &=& 2 e^{-B} (K^1 + K^3) + e^{-4 \Delta} f (K^2 - K^4) -2  m (K^2 + K^4) \nn &-&  e^{-2 \Delta -2 B}
\Re ( \mathcal{A} (T_2 - T_5).
%\label{closed1} 2 d B (S_2 + S_4) &=& e^{-4 \Delta} f \Im(L_1-L_6) - 2 m \Im(L_1+L_6) + e^{-2 \Delta -2 B} \Im(\mathcal{A}(T_4 - T_3)), \nn \\
%2 d B(S_1  - S_3) &=& -2 e^{-B} \Im(L_1-L_6) - e^{-2 \Delta - 2 B} \Im(\mathcal{A}(T_3+T_4)).
\eea
From (\ref{alg3}) and (\ref{alg4}) we get:
\bea
\label{closed2} 2 d \Delta (S_2 + S_4) &=& \tfrac{1}{4} e^{-2 \Delta - 2 B} \Im(\mathcal{A} (T_3 - T_4)) - \tfrac{1}{2} e^{-4 \Delta} f \Im(L^1-L^6) \nn &+& 2 m \Im(L^1 + L^6).
\eea
Finally, we also find useful the following torsion conditions involving scalars:
\bea
\label{scaltor1} e^{-2 \Delta} d ( e^{2 \Delta} S_1) &=& 2  S_3 d \Delta, \\
\label{scaltor2} e^{-2 \Delta} d ( e^{2 \Delta} S_3) &=& 2  S_1 d \Delta, \\
%\label{scaltor1a} d (e^{4 \Delta + B} S_2 ) &=& 2 m e^{4 \Delta} \Im(L^1) - \frac{1}{4}  e^{2 \Delta - B} \Im[\mathcal{A}(T_3 - T_4)], \\
%\label{scaltor1b} d (e^{4 \Delta + B} S_4 ) &=& 2 m e^{4 \Delta} \Im(L^6) - \frac{1}{4}  e^{2 \Delta - B} \Im[\mathcal{A}(T_3 - T_4)], \\
%\label{scaltor3} d (e^{4 \Delta +B} T_1) &=& - i e^{4 \Delta} K^1
%- 2 im e^{4 \Delta +B} K^2 + \frac{i}{4} e^{2 \Delta - B} (\mathcal{A} T_2 - \mathcal{A}^* T_5^*),  \\
%\label{scaltor4} d (e^{4 \Delta +B} T_6) &=&
%- i e^{4 \Delta} K^3 - 2 im e^{4 \Delta +B} K^4 + \frac{i}{4} e^{2 \Delta - B} (\mathcal{A}^* T_2^* - \mathcal{A} T_5),  \\
\label{scaltor5} D(e^{4 \Delta+ B} T_2) &=& \frac{i}{4} e^{2 \Delta - B } \mathcal{A}^* (T_1 - T_6^*) + \frac{i}{2} e^{4 \Delta} (1 - 2m e^{B}) (L^3 - L^4), \\
\label{scaltor6} D(e^{4 \Delta+ B} T_5) &=& \frac{i}{4} e^{2 \Delta - B } \mathcal{A}^* (T_1^* - T_6) + \frac{i}{2} e^{4 \Delta} (1 + 2m e^{B}) (L^3 - L^4).
%\label{scaltor7} D (e^{4 \Delta +B} T_3) &=& -im e^{4 \Delta +B} (L^2-L^5) - \frac{i}{2} e^{4 \Delta} (L^2 + L^5) \nn &+& \frac{i}{4} e^{2 \Delta -B}
%\mathcal{A}^*(S_1 + S_2 -S_3 + S_4), \\
%\label{scaltor8} D (e^{4 \Delta +B} T_4) &=& im e^{4 \Delta +B} (L^2-L^5) + \frac{i}{2} e^{4 \Delta} (L^2 + L^5) \nn &+& \frac{i}{4} e^{2 \Delta -B}
%\mathcal{A}^*(S_1 - S_2 -S_3 - S_4).
\eea
%Here we have grouped the results according to closely related spinor bilinears.
%In deriving these expressions we have also used (\ref{alg5}) and (\ref{alg6}) to
%rewrite them
%in terms of the warp factors.
%
%\section{$f=g=0$ torsion conditions}
In section~\ref{sec6} we employ the following bispinor relations which are valid when $f=g=0$.
%record the differential conditions satisfied
%We now determine the \textcolor{red}{some useful} differential conditions satisfied by the scalar bilinears. These may be expressed as
\bea
%\label{tors2} d S_2 &=& - m \Im(L^1) - \frac{1}{4} e^{-2 \Delta -2 B} \Im[\mathcal{A}(T_3 - T_4)], \\
%\label{tors4} d S_4 &=& - m \Im(L^6) - \frac{1}{4} e^{-2 \Delta -2 B} \Im[\mathcal{A}(T_3 - T_4)], \\
\label{t1diff} d T_1 &=& i m K^2 + \frac{i}{4}e^{-2 \Delta -2 B} [ \mathcal{A} T_2 - \mathcal{A}^* T_5^* ], \\
\label{t6diff} d T_6 &=& i m K^4 + \frac{i}{4}e^{-2 \Delta -2 B} [ \mathcal{A}^* T_2^* - \mathcal{A} T_5 ].
\eea

\end{document}